\documentclass[a4paper]{jpconf}
\usepackage{graphicx}

\def\p{\mbox{\boldmath $p$}}
\def\q{\mbox{\boldmath $q$}}
\def\beq{\begin{equation}}
\def\eeq{\end{equation}}
\def\beqn{ \begin{eqnarray} }
\def\eeqn{ \end{eqnarray} }
\newcommand{\eep} {$(e,e^{\prime}p)$ } 
\begin{document}
\title{Electron-induced proton knockout from neutron rich nuclei}

\author{C. Giusti$^1$, A. Meucci$^1$, F. D. Pacati$^1$, G. Co'$^2$, V. De
Donno$^2$,}

\address{$^1$Dipartimento di Fisica Nucleare e Teorica, 
Universit\`{a} degli Studi di Pavia and \\
INFN, Sezione di Pavia, via Bassi 6 I-27100 Pavia, Italy}

\address{$^2$Dipartimento di Fisica, Universit\`a del Salento, Lecce, and \\
INFN, Sezione di Lecce, via Arnesano, I-73100 Lecce, Italy}

\ead{Carlotta.Giusti@pv.infn.it}

\begin{abstract}
We study the evolution of the \eep cross section on nuclei with 
increasing asymmetry between the number of neutrons and protons. 
The calculations are done within the framework of the nonrelativistic and 
relativistic distorted-wave impulse approximation.  
In the nonrelativistic model phenomenological Woods-Saxon and Hartree-Fock 
wave functions are used for the proton bound-state wave functions, in the 
relativistic model the wave functions are solutions of Dirac-Hartree equations.
The models are first tested against experimental data on $^{40}$Ca and 
$^{48}$Ca nuclei, and then they are applied to a set of spherical calcium 
isotopes.
\end{abstract}

\section{Introduction}

The understanding of the evolution of nuclear properties with respect to the 
proton to neutron asymmetry is one of the major topics 
of interest in modern nuclear physics.  It is going to extend our knowledge 
about the effects of isospin asymmetry on the nuclear structure and is also 
relevant to the study of the origin and the limits of stability of matter in 
the universe. 

Nuclear reactions represent our main source of information on the
properties of atomic nuclei.  Direct 
reactions, where the external probe interacts with only one, or a few,
nucleons of the target nucleus, give insight into the
single-particle (s.p.) properties of the many-body nuclear system.  In
particular, the \eep\ reaction, where a proton is emitted
with a direct knockout mechanism, represents a clean probe to
explore the proton-hole states structure of the nucleus \cite{fru84,bof93,bof96}.
With respect to similar reactions with hadronic probes, such as, e.g., the 
$(p,2p)$ reaction, the \eep\ reaction can exploit the advantages of the 
electromagnetic interaction, that is well known and relatively weak, if 
compared with the nuclear interaction.

Several decades of experimental and theoretical work on electron scattering have
provided a wealth of information on nuclear structure and 
dynamics~\cite{bof93,bof96}. 
In the last thirty years many high-resolution exclusive \eep\ experiments 
\cite{fru84,bof93,bof96,ber82,dew90} have provided 
accurate information on the s.p. structure of stable closed-shell nuclei. 
The separation energy and the momentum distribution of the removed proton, 
which allows to determine the associated quantum numbers, have been obtained. 
From the comparison between the experimental and theoretical cross sections it 
has been possible to extract the spectroscopic factors, which give a 
measurement of the occupation of the different shells and, as a consequence, of 
the effects of nuclear correlations, which go beyond a mean field (MF) description 
of nuclear structure.

These studies can be extended to exotic nuclei.  
In upcoming years the advent of radioactive ion beams facilities
will provide a large amount of data on
unstable nuclei. A new generation of electron colliders that use
storage rings, under construction at RIKEN 
\cite{sud01} and GSI \cite{gsi06}, 
will offer unprecedented opportunities to study the structure of
exotic unstable nuclei through electron scattering in the ELISe
experiment at FAIR \cite{elise} and the SCRIT project at RIKEN \cite{sud10}. 

In this work models developed and successfully applied to the analysis of the
available experimental data for the exclusive \eep knockout reactions are
used to make predictions for the \eep cross sections on exotic nuclei. 
Our models are based on the distorted-wave impulse approximation (DWIA), within a
nonrelativistic and a relativistic framework, and are here applied to a set of
calcium isotopes. We have chosen calcium isotopes since data are available from
NIKHEF experiments for the doubly magic nuclei $^{40}$Ca and 
$^{48}$Ca \cite{kra90t,kra01}. We first compare our models with these data, 
then we apply them to some even-even isotopes, {\it i.e.}, $^{40,48,52,60}$Ca nuclei,
where the s.p. levels below the Fermi surface are fully occupied. In this 
manner we can work with spherical systems and minimize pairing effects. 
Some results are presented and discussed in the next section. More results can
be found in \cite{exot}

\section{Results}
 
Our DWIA calculations have been carried out with the same program {\tt
DWEEPY}~\cite{giu87,giu88} which was used for the analyses of the 
NIKHEF data. The program includes the effects of the final-state
interactions (FSI) between the emitted proton and the remaining nucleus, that 
are described in the calculations by the phenomenological optical potential of 
\cite{sch82}, as well as of the Coulomb distortion of the electron
wave functions. Although based on a nonrelativistic DWIA model, {\tt DWEEPY} 
contains relativistic effects in the kinematics and in the nuclear
current operator.
In the comparison with the $^{40}$Ca\eep and  $^{48}$Ca\eep data we have 
repeated the original analyses of \cite{kra90t,kra01}, where
the bound-state wave functions are calculated with a phenomenological 
Woods-Saxon (WS)
well whose radius was determined to fit the experimental momentum
distribution and whose depth was adjusted to give the experimentally
observed separation energy of the bound final state. This choice was able to 
describe, with a high degree of accuracy, the shape of the experimental 
momentum distributions at missing-energy values corresponding to specific 
peaks in the energy spectrum. In order to reproduce the size of the 
experimental cross sections, 
reduction factors must be applied to the calculated cross sections. These 
factors are identified with the spectroscopic factors and their deviation   
from the predictions of the MF approximation is interpreted as the effect of 
nucleon-nucleon correlations.

The results obtained with  WS wave functions are here compared 
with those obtained by solving Hartree-Fock (HF) equations with the Gogny-like
finite-range D1M interaction \cite{gor09}. 

In a third approach we have performed \eep calculations with the relativistic 
DWIA (RDWIA) model of \cite{meu01a,meu01,meu02a,meu02b,radici03,tamae09}, 
where the s.p. bound-state wave function is the Dirac-Hartree
solution from a relativistic Lagrangian written in the context of the 
relativistic MF theory (RMF) \cite{hor91} and the scattering wave function is solution of 
the Dirac equation with the relativistic energy-dependent and A-dependent EDAD1
optical potential \cite{coo93}.

\begin{figure}
\begin{center}
\includegraphics[width=100mm, height=60mm]{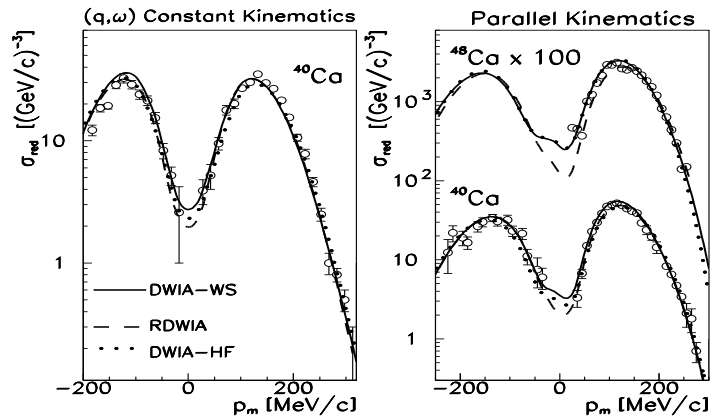}
\end{center}
\vskip  -3mm
\caption{Reduced cross sections of the
  $^{40}$Ca\eep and $^{48}$Ca\eep reactions as a function of 
  $p_{\mathrm m}$ for the transition to the $3/2^{+}$ ground state of $^{39}$K
  and to the $3/2^{+}$ excited state at 0.36 MeV of $^{47}$K in ($\q,\omega$) 
  constant kinematics (left panel), with incident electron energy $E_0= 483.2$ 
  MeV, electron scattering angle $\vartheta= 61.52^\circ$, and $q=450$ MeV/$c$,  
  and in parallel kinematics (right panel), with $E_0= 483.2$ MeV. 
  The outgoing proton energy is $T'= 100$ MeV in both kinematics. 
  Line convention: DWIA-WS (solid lines), DWIA-HF (dotted lines), RDWIA (dashed
  line).
%Positive (negative) values of $p_{\mathrm m}$
%refer to situations where in ($\q,\omega$) constant kinematics the
%angle between the outgoing proton momentum $\p'$ and the incident
%electron $\p_0$ is larger (smaller) than the angle between $\q$ and
%$\p_0$. In parallel kinematics positive and negative values of
%$p_{\mathrm m}$ indicate, respectively, the $|\q|<|\p'|$ and
%$|\q|>|\p'|$ cases. 
  The experimental data are taken from \cite{kra90t}.  
\label{fig1}}
\end{figure}

In Fig. \ref{fig1} our DWIA and RDWIA results are compared with the
$^{40}$Ca\eep and $^{48}$Ca\eep NIKHEF data \cite{kra90t} for the knockout of a
proton from the  $1d_{3/2}$ s.p. level. 
Data for the $^{40}$Ca\eep reaction were taken in the so-called parallel and 
($\q,\omega$) constant kinematics. For the $^{48}$Ca\eep 
reaction data were taken only in parallel kinematics. 
In parallel kinematics the momentum of the outgoing proton $\p'$ is kept fixed 
and is taken parallel to the momentum transfer $\q$. Different 
values of the missing momentum $p_{\mathrm m}$, which is the recoil momentum of
the residual nucleus, are obtained by varying the 
electron scattering angle and, as a consequence, $q$. 
In ($\q,\omega$) constant kinematics $\q$ and the outgoing proton energy are 
kept constant and different values of $p_{\mathrm m}$ are obtained by varying 
the angle of the outgoing proton.
Experimental data are usually presented as a function of $p_{\mathrm m}$ in 
terms of the reduced cross section \cite{bof96}, {\it i.e.}, the cross section 
divided by a kinematical factor and by the elementary off-shell 
electron-proton scattering cross section of \cite{def83}.
In the plane-wave impulse approximation (PWIA), where FSI are neglected and  
the cross section is factorized into the product of a kinematical factor, the 
elementary electron-proton scattering cross section, and the hole spectral 
function, the reduced cross section is the squared 
Fourier transform of the hole wave function, and can be interpreted as the 
momentum distribution of the emitted proton when it was inside the nucleus.
In DWIA this factorization is destroyed by FSI, but the reduced cross section 
remains an interesting quantity that can be regarded as the nucleon momentum 
distribution modified by FSI.

All the theoretical results provide a good
description of the shape of the experimental data. A reduction factor has been
applied to reproduce the size. 
This factor has been determined by a fit of the calculated reduced cross 
sections to the data over the whole missing-momentum range considered in the
experiment. The reduction factors applied to the 
DWIA-WS results (0.49 in ($\q,\omega$) constant 
kinematics and 0.65 in parallel kinematics for $^{40}$Ca, 0.55 for $^{48}$Ca) 
are the same as in the data analysis of \cite{kra90t}. The
corresponding factors for the DWIA-HF results are 0.51, 0.64, 0.55, and for the  
RDWIA results are 0.49, 0.69, 0.52.  
In all the calculations the reduction factors obtained for $^{40}$Ca\eep
reaction  in ($\q,\omega$) constant kinematics are about 20-25\%
lower than those obtained in parallel kinematics. The source of this difference is not 
clear, but it reflects the uncertainties in the identification of the 
spectroscopic factor as a simple reduction factor of the theoretical results 
with respect to the experimental data. The reduction factors obtained in all our
model calculations for the $^{48}$Ca\eep reaction are consistently lower than 
those obtained for the $^{40}$Ca\eep reaction in the same parallel kinematics.

\begin{figure}
\begin{center}
\includegraphics[width=100mm, height=90mm]{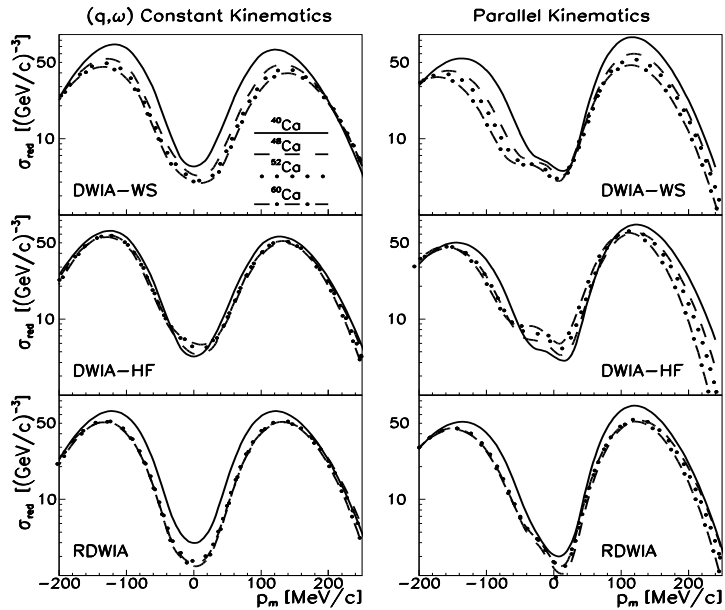}
\end{center}
\vskip  -3mm
\caption{Reduced cross section of the \eep reaction for
$1d_{3/2}$ knockout from
$^{40}$Ca (solid lines), $^{48}$Ca (dashed lines), $^{52}$Ca (dotted
lines), and $^{60}$Ca (dot-dashed lines), as a function of $p_{\mathrm m}$,
calculated in ($\q,\omega$) constant (left panels) and parallel kinematics
(right panels) with DWIA-WS (top panels), DWIA-HF (middle panels), and RDWIA 
(bottom panels). 
%The results in the left panels are obtained in ($\q,\omega$) constant 
%kinematics, with $E_0= 483.2$ MeV, $\vartheta= 61.52^{\circ}$, and $q=450$ MeV/$c$. 
%The results in the right panels are obtained in parallel kinematics 
%with $E_0= 440$ MeV and $T'= 100$ MeV.
\label{fig2}}
\end{figure}
The predictions of our models for $1d_{3/2}$ knockout from
$^{40,48,52,60}$Ca nuclei in ($\q,\omega$) constant and parallel kinematics 
are shown in Fig. \ref{fig2}.   
We remark that the evolution of the cross
section with respect to the change of the neutron number is
the same in all the panels: the
$^{40}$Ca lines are always above the other ones, and the size of
the curves decreases with the increasing number of neutrons. This
behavior is clearer in the DWIA-WS results and becomes less evident in the 
other cases, especially in the RDWIA ones.
The behavior of the s.p. hole wave functions with increasing neutron number
shows a different trend for the different models \cite{exot}. 
The dependence of the wave functions on the  proton to neutron 
asymmetry is responsible for only a part of the differences in the reduced 
cross sections. While in PWIA the reduced cross section contains only 
information on the bound-state wave function, in DWIA this information is 
modified by the contribution of the 
other ingredients of the model, such as FSI and the electron-nucleon 
interaction. All these contributions are intertwined in the calculated cross 
section and, in general, they cannot be easily disentangled.
The difference between the results of the two kinematics in 
Fig.~\ref{fig2} is basically due to the different effects of the
distortion produced by the optical potential. These effects strongly
depend on kinematics and are larger in parallel than in ($\q,\omega$)
constant kinematics. 
Moreover, the trend of the calculated cross sections with the increasing neutron
number is significanlty affected by the A-dependence of the optical potential
\cite{exot}. 

No reduction factor has been applied to the results shown in Fig.~\ref{fig2}.
The comparison with data in Fig.~\ref{fig1} gives a significant 
quenching of the measured cross sections with respect to the predictions of 
the MF model. The quenching is different for the $^{40}$Ca\eep\ and
$^{48}$Ca\eep\ reactions and increases with the neutron number. A
quenching depending on the neutron number can be expected for all
the isotopes and would give further differences on the reduced cross
sections than those shown in Fig.~\ref{fig2}.

\section{Summary and conclusions}

We have presented and discussed \eep cross sections
for a set of calcium isotopes with the aim of
studying their evolution with respect to the change of the neutron
number.  

The nonrelativistic DWIA and relativistic RDWIA models used for the 
calculations were widely and successfully applied to the analysis of the 
available \eep data over a wide range of stable nuclei.  The results obtained 
with three different MF descriptions of the hole wave function of the knocked
out proton have been compared.  

All the three models give a good and similar description of the 
experimental data on $^{40}$Ca and $^{48}$Ca. 
The general behavior of the cross sections with respect to the  
increasing number of neutrons is analogous for all the three
models: the reduced cross sections are larger and
narrower for the lighter isotopes, and evolve by lowering and widening
with increasing neutron number.
Although the evolution of the s.p. bound-state wave functions is different for 
the three models, 
the dependence of the wave functions on the proton to neutron 
asymmetry is responsible for only a part of the
differences in the calculated cross sections. An important and crucial
contribution is also given by FSI, which are described in our models by 
phenomenological optical potentials. The optical
potential affects both
the size and the shape of the cross section in a way that strongly
depends on kinematics.  In particular, its imaginary part, that  gives a 
reduction of the calculated cross 
section, can affect the values of the spectroscopic factors obtained from the 
comparison between data and theoretical results.
The dependence of the optical potential on the proton to neutron 
asymmetry is an interesting problem that deserves careful investigation.

From recent studies there are indications
that the spectroscopic factors and the effects of correlations depend
on the proton to neutron asymmetry. In general, the
quenching of quasiparticle orbits, and hence correlations, become
stronger with increasing separation energy.

Measurements of the exclusive quasifree \eep cross section on nuclei
with neutron excess would offer a unique opportunity for studying the
dependence of the properties of bound protons on the neutron to proton 
asymmetry.
The present results can serve as a useful reference for future experiments. 
The comparison with data can confirm or invalidate the predictions of the model
and will test the ability of the established nuclear theory in the domain of
exotic nuclei.

\vskip  5mm

This work was partially supported by the MIUR through the PRIN 2009 research project.

\end{document}